\RequirePackage{fix-cm} 
\documentclass[a4paper,twoside,reqno,dvips,12pt]{amsart}
\usepackage{fixltx2e}     

\usepackage[latin1]{inputenc}
\usepackage[T1]{fontenc}

\usepackage{eucal,verbatim}
\usepackage{esint}
\usepackage{xspace}
\usepackage{amsgen}
\usepackage{amssymb}
\usepackage{amsmath}
\usepackage{amsfonts}
\usepackage{MnSymbol}
\usepackage[nice]{nicefrac}
\usepackage{mathrsfs, dsfont}

\usepackage{a4wide}

\usepackage[table]{xcolor}

\definecolor{oneblue}{rgb}{0.0, 0.0, 0.85}
\definecolor{darkgrey}{rgb}{0.273, 0.281, 0.30}
\definecolor{Lightgray}{rgb}{0.89, 0.89, 0.89}
\definecolor{Lightblue}{RGB}{214, 214, 214}
\definecolor{bckg}{RGB}{20.8, 20.8, 20.8} 

\usepackage{psfrag}
\usepackage{graphicx}
\usepackage{subfigure}
\usepackage{morefloats}
\usepackage{indentfirst}

\usepackage{acronym}
\usepackage{microtype}
\usepackage[labelfont={bf,sf},%
            labelsep=period,%
            justification=raggedright]{caption}

\usepackage[usenames, dvipsnames, pdf]{pstricks}
\usepackage{epsfig}
\usepackage{pst-grad} 
\usepackage{pst-plot} 

\usepackage[colorlinks,
            urlcolor=oneblue,
            linkcolor=oneblue,
            citecolor=oneblue,
            bookmarksopen=false,
            pagebackref]{hyperref}

\usepackage[explicit]{titlesec}

\titleformat{\section}
  {\color{darkgrey}\Large\sffamily\bfseries}
  {}
  {0em}
  {\colorbox{bckg!5}{\parbox{\dimexpr\linewidth-2\fboxsep\relax}{\centering\thesection. #1}}}
  []

\titleformat{name=\section,numberless}
  {\color{darkgrey}\Large\sffamily\bfseries}
  {}
  {0em}
  {\colorbox{bckg!10}{\parbox{\dimexpr\linewidth-2\fboxsep\relax}{\centering#1}}}
  []

\titleformat{\subsection}
  {\color{darkgrey}\large\sffamily\bfseries}
  {}
  {0em}
  {\colorbox{bckg!5}{\parbox{\dimexpr\linewidth-2\fboxsep\relax}{\centering\thesubsection. #1}}}
  []

\titleformat{name=\subsection,numberless}
  {\color{darkgrey}\Large\sffamily\bfseries}
  {}
  {0em}
  {\colorbox{bckg!10}{\parbox{\dimexpr\linewidth-2\fboxsep\relax}{\centering#1}}}
  []

\titleformat{\subsubsection}
  {\sffamily\normalsize\bfseries}
  {\thesubsubsection}
  {0em}
  {#1}
  []

\titleformat{\paragraph}[runin]
  {\sffamily\small\bfseries}
  {}
  {0em}
  {#1}

\titlespacing*{\section}{1.0em}{1.0em}{0.8em}[0em]
\titlespacing*{\subsection}{1.0em}{1.0em}{0.8em}[0em]
\titlespacing*{\subsubsection}{1.0em}{0.7em}{0.6em}[0em]

\usepackage{titletoc}

\contentsmargin{0.55em}

\newlength{\tocsep}
\titlecontents{section}[\tocsep]
  {\addvspace{10pt}\bfseries\sffamily}
  {\contentslabel[\thecontentslabel]{\tocsep}}
  {}
  {\ \titlerule*[0.75pc]{.}\ \thecontentspage}
  []
\titlecontents{subsection}[\tocsep]
  {\addvspace{8pt}\sffamily}
  {\contentslabel[\thecontentslabel]{\tocsep}}
  {}
  {\ \titlerule*[0.5pc]{.}\ \thecontentspage}
  []
\titlecontents{subsubsection}[\tocsep]
  {\addvspace{7pt}\footnotesize\sffamily}
  {\contentslabel[\thecontentslabel]{\tocsep}}
  {}
  {\ \titlerule*[0.5pc]{.}\ \thecontentspage}
  []

\usepackage{fancyhdr}
\usepackage{lastpage}

\newcommand*\Title{Plethora of generalised solitary waves}
\newcommand*\Authors{D.~Clamond, D.~Dutykh \& A.~Dur\'an}
\newcommand*{\plogo}{{\texttt{arXiv.org} / \textsc{hal}}} 

\numberwithin{equation}{section}

\newcommand{\sur}[1]{{#1}_{\mathrm{s}}}                   
\newcommand{\bott}[1]{{#1}_{\mathrm{b}}}                  

\newcommand{\depth}{d}

\newcommand{\ud}{\mathrm{d}}
\newcommand{\ui}{\mathrm{i}}
\newcommand{\ue}{\mathrm{e}}

\newcommand{\Bo}{\mathsf{Bo}}
\newcommand{\Fr}{\mathsf{Fr}}

\renewcommand{\Re}{\operatorname{Re}}
\renewcommand{\Im}{\operatorname{Im}}

\newcommand{\half}{{\textstyle{1\over2}}}

\newcommand{\halfi}{{\textstyle{1\over2\ui}}}

\newcommand{\eg}{e.g.}

\acrodef{lm}[LM]{Levenberg--Marquardt}

\begin{document}

\title[\Title]{A plethora of generalised solitary \\ gravity-capillary water waves}

\author[D. Clamond]{Didier Clamond$^*$}
\address{Universit\'e de Nice -- Sophia Antipolis, Laboratoire J. A. Dieudonn\'e, Parc Valrose, 06108 Nice cedex 2, France}
\email{diderc@unice.fr}
\urladdr{http://math.unice.fr/~didierc/}
\thanks{$^*$ Corresponding author}

\author[D. Dutykh]{Denys Dutykh}
\address{Universit\'e Savoie Mont Blanc, LAMA, UMR 5127 CNRS, Campus Scientifique, 73376 Le Bourget-du-Lac Cedex, France}
\email{Denys.Dutykh@univ-savoie.fr}
\urladdr{http://www.denys-dutykh.com/}

\author[A.~Dur\'an]{Angel Dur\'an}
\address{Departamento de Matem\'atica Aplicada, E.T.S.I. Telecomunicaci\'on, Campus Miguel Delibes, Universidad de Valladolid, Paseo de Belen 15, 47011 Valladolid, Spain}
\email{angel@mac.uva.es}


\begin{titlepage}
\setcounter{page}{1}
\thispagestyle{empty} 
\noindent
{\Large Didier \textsc{Clamond}}\\
{\it\textcolor{gray}{Universit\'e de Nice -- Sophia Antipolis, France}}\\[0.02\textheight]
{\Large Denys \textsc{Dutykh}}\\
{\it\textcolor{gray}{Universit\'e Savoie Mont Blanc, France}}\\[0.02\textheight]
{\Large Angel \textsc{Dur\'an}}\\
{\it\textcolor{gray}{Universidad de Valladolid, Spain}}\\[0.16\textheight]

\colorbox{Lightblue}{
  \parbox[t]{1.0\textwidth}{
    \centering\huge\sc
    \vspace*{0.7cm}

    A plethora of generalised solitary \\ gravity-capillary water waves

    \vspace*{0.7cm}
  }
}

\vfill 

\raggedleft     
{\large \plogo} 
\end{titlepage}


\begin{abstract}
The present study describes, first, an efficient algorithm for computing capillary-gravity solitary waves solutions of the irrotational Euler equations with a free surface and, second, provides numerical evidences of the existence of an infinite number of generalised solitary waves (solitary waves with {\em undamped\/} oscillatory wings). Using conformal mapping, the unknown fluid domain, which is to be determined, is mapped into a uniform strip of the complex plane. In the transformed domain, a Babenko-like equation is then derived and solved numerically.

\bigskip
\noindent \textbf{\keywordsname:} Solitary surface waves; capillary--gravity waves; Euler equations; generalised solitary waves.


\end{abstract}

\newpage
\tableofcontents
\thispagestyle{empty}
\newpage


\section{Introduction}

Despite numerous studies devoted to capil\-lary--gravity waves, this topic still fascinates the researchers. The review by \cite{Dias1999} shortly summarises in great lines what has been known on this subject at the end of the twentieth century. Monographs by \cite{Okamoto2001} and by \cite{Vanden-Broeck2010} are other exhaustive sources of informations on various types of capillary--gravity waves.

The travelling capillary--gravity waves of permanent form have been most deeply understood in the framework of weakly nonlinear and weakly dispersive equations, such as the Korteweg--de Vries (KdV) equation and the extended Korteweg--de Vries with fifth-order derivatives (KdV5) equation. These equations model the unidirectional propagation of long waves in shallow water with some weak capillarity. Despite the apparent simplicity of the KdV5 model, it possesses a rich family of solutions. One of them consists of the so-called {\em generalised solitary waves}. These are solitary-wave pulses that are homoclinic to small amplitude oscillatory waves. The formation of these waves is mathematically justified by the presence of a resonance  in the dynamical system corresponding to the traveling waves \cite{Lombardi2000}. The existence of generalised solitary waves can be deduced from the existence of the phase shift when the action of the stationary problem changes sign \cite{Bridges2005}. One can also use the existence of Smale's horseshoe dynamics on the zero energy set \cite{Buffoni1996}. Then, it is relatively straightforward to construct a symbolic orbit which represents a generalised solitary wave. By using the first method one obtains a continuum family of solutions, while the latter gives `only' a countable set of orbits. \cite{Benilov1993} showed that, for sufficiently weak surface tensions, there cannot exist generalised solitary waves. For non-vanishing values of the surface tension \cite{Grimshaw1995} constructed a one-parameter family of generalised solitary waves for the KdV5 equation using the methods of exponential asymptotic, see also \cite{YangAkylas1996}. The existence of multi-pulse solutions was shown numerically and analytically \cite{CalvoAkylas1997,Champneys1997}. The main difficulties to compute numerically  the generalised solitary waves for the KdV5 equation are well described by \cite{Boyd2007}. The stability of multi-pulse solitary waves was studied by \cite{Chardard2009} using the Maslov theory.

Much less is known for the full water wave problem. However, the existence of generalised capillary--gravity solitary waves for the full Euler equations was shown by \cite{Sun1991} and \cite{Sun1993}. As highlighted by \cite{Beale1991}, the generalised solitary waves stem from a resonance with periodic waves of the same speed. By using the method of boundary integral equations and Newton iterations to solve the resulting discrete system, \cite{Hunter1983a} computed solitary and periodic capillary-gravity waves in finite depth (see also \cite{Vanden-Broeck2010} and the references therein) while multi-modal solitary waves of depression are generated in \cite{Dias1996} in both finite and infinite depth. Recently, generalised solitary waves were computed to the full water wave problem \cite{Champneys2002}. The main purpose of the present paper is to show that there is a plethora (likely an infinite number) of generalised solitary waves for the full Euler equations.

In the present study, we consider a formulation for steady capillary--gravity solitary waves by following the pioneering work of \cite{Babenko1987}. This formulation is based on the conformal mapping technique \cite{Dyachenko1996a, Dyachenko1996, Ovsyannikov1974} that has been successfully used to compute numerically periodic gravity waves in deep water \cite{Choi1999} and in finite depth \cite{Li2004, Longuet-Higgins1996}. More recently, this approach was adapted also to periodic capillary--gravity waves in deep water \cite{Milewski2010}. The advantage of the Babenko's formulation is that it does not add nonlinearities to the Euler equations in the conformally mapped domain. For instance, the Babenko equation being quadratic in nonlinearity for pure gravity waves, it is easily solved numerically for solitary waves \cite{Clamond2012b}. This approach is extended here to compute generalised solitary waves.

In order to solve numerically the Babenko equation for capillary--gravity solitary waves, it is discretised using the Fourier-type pseudo-spectral method \cite{Boyd2000}. The resulting system of nonlinear equations is solved using the well-known \acf{lm} method \cite{Levenberg1944, Marquardt1963}. This algorithm represents a mixture between the steepest descent far from the solution and the classical Newton method in the vicinity of the solution \cite{More1978}. It has been shown to be a robust nonlinear solver even in problems with millions of unknowns \cite{Lourakis2005}.

The paper is organised as follows. In Section~\ref{sec:model}, we present the main constitutive assumptions of the mathematical model, as well as the conformal mapping technique and the derivation of a Babenko-like equation. The numerical resolution procedure is explained in Section~\ref{sec:numm}. Numerical results are presented in Section~\ref{sec:numr}. The new results on multi-hump generalised solitary waves are discussed in Section~\ref{sec:concl}.


\section{Mathematical model}\label{sec:model}

We consider a steady two-dimensional potential flow induced by a solitary wave in a horizontal channel of constant depth. The fluid is assumed to be inviscid and homogeneous. The pressure is equal to the surface tension at the impermeable free surface and the fixed horizontal seabed is impermeable as well. The flow is driven by the volumetric gravity force (directed downward) and by the capillary forces at the free surface.

Let $(x,y)$ be a Cartesian coordinate system moving with the wave, $x$ being the horizontal coordinate and $y$ being the upward vertical one. The equations $y=-\depth$, $y=\eta(x)$ and $y=0$ denote the positions of the bottom, of the free surface and of the mean water level, respectively. The latter implies that the Eulerian average $\left<\bullet\right>$ of the free surface is zero: 
\begin{equation}\label{defmean}
  \left<\,\eta\,\right>\ \equiv\ \lim_{L\to\infty}{1\over2\,L}\int_{-L}^{L}\eta(x)\ \ud\/x\ =\ 0.
\end{equation}
This definition of the average operator is not suitable for the computation of solitary waves. However, for a classical solitary wave, the mean condition (\ref{defmean}) implies that 
\begin{equation}\label{defmeansolcla}
  \eta(x)\ \to\ 0 \qquad\text{as}\qquad x\ \to\ \pm\infty,
\end{equation}
the latter being more tractable for computations (see Section \ref{sec:numm}). Also, for classical solitary waves, $|\eta|$ decaying faster than $1/|x|$ in the far field, the condition (\ref{defmean}) implies that the wave mass $\int_{-\infty}^\infty\eta(x)\,\ud\/x$ is finite.Here, we consider finite mass generalised solitary waves with a $(2\ell)$-periodic tail in the far field. For such waves, the condition (\ref{defmean}) yields
\begin{equation}\label{defmeansolgen}
  \frac{1}{2\,\ell}\int_{x-\ell}^{x+\ell}\eta(\xi)\,\ud\/\xi\ \to\ 0 \qquad\text{as}\qquad x\ \to\ \pm\infty.
\end{equation}
The definition (\ref{defmeansolgen}) of the mean level is more suitable than (\ref{defmean}) for practical computations of generalised solitary waves (see Section \ref{sec:numm}). Obviously, the condition (\ref{defmeansolgen}) is equivalent to (\ref{defmeansolcla}) when the amplitude of the tail tends to zero.

Let $\phi$, $\psi$, $u$ and $v$ be, respectively, the velocity potential, the stream function, the horizontal and vertical velocities, such that $u\equiv\partial_x\phi=\partial_y\psi$ and $v\equiv\partial_y\phi=-\partial_x\psi$. It is convenient to introduce the complex potential $f \equiv \phi + \ui\/\psi$ (with $\ui^2 = -1$) and the complex velocity $w \equiv u - \ui\/v$ that are holomorphic functions of $z\equiv x+\ui\/y$ ($w = \ud f/\ud z$). The complex conjugate is denoted with a star (\eg, $z^* = x - \ui\/y$), while subscripts `b' denote the quantities evaluated at the seabed --- \eg, $\bott{z}(x)=x-\ui\depth$, $\bott{\phi}(x)=\phi(x,y\!=\!-\depth)$ --- and subscripts `s' denote the quantities evaluated at the surface --- \eg, $\sur{z}(x) = x + \ui\eta(x)$, $\sur{\phi}(x) = \phi(x,y\! = \!\eta(x))$. Note that, \eg, $\sur{u} = \sur{(\partial_x\phi)} \neq\partial_x \sur{\phi} = \sur{u} + \eta_x\sur{v}$. The traces of $\psi$ on the upper and lower boundaries, $\sur{\psi}$ and $\bott{\psi}$, are constants because the surface and the bottom are streamlines. These constants are related to the mean flow by
\begin{equation}\label{defc}
  -\,c\ \equiv\ \frac{1}{\depth}\left<\,\int_{-\depth}^\eta u\,\ud\/y\,\right>\,=\ \frac{\sur{\psi}\,-\,\bott{\psi}}{d},
\end{equation}
so $c$ is the wave phase velocity observed in the frame of reference without mean flow ($c>0$ if the wave travels toward the increasing $x$-direction, so we consider $c>0$ without loss of generality).

The dynamic condition can be expressed in term of the Bernoulli equation
\begin{equation}\label{bernbase}
  2\,p\ +\ 2\,g\,y\ +\ u^2\ +\ v^2\ =\ B,
\end{equation}
where $p$ is the pressure divided by the density, $g > 0$ is the acceleration due to gravity and $B$ is a Bernoulli constant. At the free surface $y=\eta(x)$ the pressure $p$ reduces to the effect of the surface tension, that is $\sur{p}=-\tau\,\partial_x\!\left[\eta_{x}\,(1+\eta_x^{\,2})^{-1/2}\right]$, $\tau$ being a (constant) surface tension coefficient (divided by the density). Averaging \eqref{bernbase} written at the free surface, the definition of the mean level \eqref{defmean} yields an equation for the Bernoulli constant $B$
\begin{equation*}
  B\ =\,\left<\,\sur{u}^{\,2}\, + \,\sur{v}^{\,2}\,\right>.
\end{equation*}

In the far field of a classical solitary waves, the free surface is horizontal and the flow is an uniform current. Thus, $\eta\to0$, $u\to-c$ and $v\to0$ as $x\to\pm\infty$, where $c$ is defined in (\ref{defc}). It follows that $B=c^2$ for classical solitary waves. 

For the generalised solitary waves considered here, the far field involves a periodic wave of constant amplitude and it is therefore not an uniform current, implying that $B\neq c^2$ in general. By analogy of the classical solitary wave, we define another phase velocity by
\begin{equation}\label{defcspeed}
  c'\ \equiv\ \sqrt{\/B\/}.
\end{equation}
Many other phase velocities could of course be defined, such as the mean horizontal velocity at the bottom in the far field, but these considerations are secondary for the purpose of the present work. Note that the definition (\ref{defcspeed}) implies that $c'>0$ without loss of generality.


\subsection{Conformal mapping}\label{sec:conmap}

Let be the change of independent variable  $z \mapsto \zeta \equiv (\ui\sur{\psi} - f)/c$, that conformally maps the fluid domain
\begin{equation}
  -\infty\ \leqslant\ x\ \leqslant\ \infty, \qquad
  -\,\depth\ \leqslant\ y\ \leqslant\ \eta(x),
\end{equation}
into the strip
\begin{equation}
  -\infty\ \leqslant\ \alpha\ \leqslant\ \infty, \qquad
  -\,\depth\ \leqslant\ \beta\ \leqslant\ 0,
\end{equation}
where $\alpha \equiv \Re(\zeta)$ and $\beta \equiv \Im(\zeta)$. The conformal mapping yields the Cauchy--Riemann relations $x_\alpha = y_\beta$, $x_\beta = -y_\alpha$ thence we have 
\begin{eqnarray} \label{wuvqmap}
  \frac{c}{w}\/ =\/ -\frac{\ud\,z}{\ud\/\zeta}, \qquad
  \frac{u}{c}\/ =\/ \frac{-\,x_\alpha}{x_\alpha^{\,2}\/+\/y_\alpha^{\,2}}, \qquad
  \frac{v}{c}\/ =\/ \frac{-\,y_\alpha}{x_\alpha^{\,2}\/+\/y_\alpha^{\,2}}, \qquad
  \frac{u^2\/+\/v^2}{c^2}\/ =\/ \frac{1}{x_\alpha^{\,2}\/+\/y_\alpha^{\,2}},
\end{eqnarray}
while the pressure at the free surface can be conveniently written (see Appendix \ref{appmath})
\begin{equation}\label{psdzs}
  \frac{\sur{p}}{\tau}\ =\ -\left(\frac{\ud\,\sur{z}^*}{\ud\/\alpha}\right)^{\!-1}\/\frac{\ud}{\ud\/\alpha}\!\left[\frac{\ui\,x_\alpha\,+\,y_\alpha}{\sqrt{\,x_\alpha^{\,2}\,+\,y_\alpha^{\,2}\,}}\right]_{\!\beta=0}.
\end{equation}

With the change of dependent variables
\begin{eqnarray}\label{eq:X}
  x\ =\ \alpha\ +\ X(\alpha,\beta), \qquad
  y\ =\ \beta\ +\ Y(\alpha,\beta),
\end{eqnarray}
the Cauchy--Riemann relations $X_\alpha = Y_\beta$ and $X_\beta = -Y_\alpha$ hold, while the bottom ($\beta = -\depth$) and the free surface ($\beta = 0$) impermeabilities yield
\begin{eqnarray}\label{Ybotsur}
  \bott{Y}(\alpha)\ \equiv\ Y(\alpha,-\depth)\ =\ 0, \qquad
  \sur{Y}(\alpha)\ \equiv\ Y(\alpha,0)\ =\ \eta(\alpha).
\end{eqnarray}

The functions $X$ and $Y$ can be expressed in terms of $\bott{X}$ as \cite{Clamond1999, Clamond2003}
\begin{align}\label{eq:Xbot}
  X(\alpha,\beta)\ &=\ \half\,\bott{X}(\zeta+\ui\/\depth)\ +\ \half\,\bott{X}(\zeta^*-\ui\/\depth)\ 
  =\ \cos\!\left[\,(\beta+\depth)\/\partial_\alpha\,\right]\bott{X}(\alpha), \\
  Y(\alpha,\beta)\ &=\ \halfi\,\bott{X}(\zeta+\ui\/\depth)\ -\ \halfi\,\bott{X}(\zeta^*-\ui\/\depth)\ 
  =\ \sin\!\left[\,(\beta+\depth)\/\partial_\alpha\,\right]\bott{X}(\alpha),\label{eq:Ybot}
\end{align}
where a star denotes the complex conjugate. Thus, the Cauchy--Riemann relations and the bottom impermeability are fulfilled identically. At the free surface $\beta=0$, \eqref{eq:Xbot} yields
\begin{equation*}
  \sur{X}(\alpha)\ =\ \cos\!\left[\,\depth\,\partial_\alpha\,\right]\bott{X}(\alpha)
\qquad\Longleftrightarrow\qquad
  \bott{X}(\alpha)\ =\ \sec\!\left[\,\depth\,\partial_\alpha\,\right]\sur{X}(\alpha),
\end{equation*}
and hence the relation \eqref{eq:Ybot} yields
\begin{equation}\label{eq:relYXs}
  \sur{Y}(\alpha)\ =\ \tan\!\left[\,\depth\,\partial_\alpha\,\right]
  \sur{X}(\alpha)\ \equiv\ \mathscr{T}\!\left\{\sur{X}\right\},
\end{equation}
which relates quantities written at the free surface only. The relation \eqref{eq:relYXs} can be trivially inverted giving, in particular,
\begin{equation}\label{eq:relYXs2}
  \sur{(\partial_\alpha X)}\ =\ \partial_\alpha\cot\!\left[\,\depth\,\partial_\alpha\,\right]\,
  \sur{Y}\ \equiv\ \mathscr{C}\{\sur{Y}\}\ =\ \mathscr{C}\{\eta\},
\end{equation}
where $\mathscr{T}$ and $\mathscr{C}$ are  pseudo-differential operators that, for a pure frequency, take the form
\begin{eqnarray}\label{mathC}
  \mathscr{T}\!\left\{\ue^{\ui k\alpha}\right\} = \ui\tanh(k\depth)\,\ue^{\ui k\alpha}, \quad
  \mathscr{C}\!\left\{\ue^{\ui k\alpha}\right\}\,=\,\left\{\begin{array}{lr}
  k\coth(k\depth)\,\ue^{\ui k\alpha}  &\quad  (k\neq0),  \\
  1\,/\,\depth  & \quad  (k=0).
\end{array}\right. \qquad
\end{eqnarray}


\subsection{Babenko-like equation}

Using (\ref{wuvqmap}{\it a}), (\ref{psdzs}) and the relation $u^2+v^2=w\/w^*$, the Bernoulli equation (\ref{bernbase}) at the free surface can be written
\begin{align}\label{defwsur}
  \sur{w}\ &=\ \frac{B\,-\,2\,g\,\eta\,-\,2\,\sur{p}}{\sur{w}^*}\ =\ \frac{2\,\sur{p}\,+\,2\,g\,\eta\,-\,B}{c}\,\frac{\ud\,\sur{z}^*}{\ud\/\alpha} \nonumber\\
  &=\ \frac{2\,g\,\eta\,-\,B}{c}\,\frac{\ud\,(\sur{x}-\ui\eta)}{\ud\/\alpha}\ -\ \frac{2\,\tau}{c}\,\frac{\ud}{\ud\/\alpha}\!\left[\frac{\ui\,x_\alpha\,+\,y_\alpha}{\sqrt{\,x_\alpha^{\,2}\,+\,y_\alpha^{\,2}\,}}\right]_{\!\beta=0}.
\end{align}
$w(\zeta)=u(\alpha,\beta)-\ui\/v(\alpha,\beta)$ being a holomorphic function such that $\Im(w)=0$ at the bottom, we have at the free surface --- see above the derivation of (\ref{eq:relYXs}) ---
\begin{equation}\label{relRSsur}
  -\,\sur{v}(\alpha)\ =\ \mathscr{T}\!\left\{\sur{u}\right\}\ =\ \tan[\,\depth\,\partial_\alpha\,]\,\sur{u}(\alpha).
\end{equation}
Applying the operator $\cot[\/\depth\,\partial_\alpha\/]=\partial_\alpha^{\,-1}\,\mathscr{C}$ to (\ref{relRSsur}), then exploiting the relations (\ref{eq:X}), (\ref{eq:relYXs2}) and (\ref{defwsur}), after some algebra we obtain a pseudo-differential equation for $\eta$ 
\begin{align}\label{eq:bab1}
  &\mathscr{C}\!\left\{B\,\eta\,-\,\frac{g\,\eta^2}{2}\,+\, \tau\,-\,\frac{\tau\,(\/1\/ + \/\mathscr{C}\{\eta\}\/)}{\sqrt{\/(1\/+\/\mathscr{C}\{\eta\})^2\/+\/\eta_\alpha^{\,2}\/}} \right\} \nonumber\\ &-\ g\,\eta\,(1\/+\/\mathscr{C}\{\eta\})\ +\ \frac{\ud}{\ud\/\alpha}\left\{\frac{\tau\,\eta_\alpha}{\sqrt{\/(1\/+\/\mathscr{C}\{\eta\})^2\/+\/\eta_\alpha^{\,2}\/}}\right\}\,=\ K,
\end{align}
where $K$ is an integration constant to be determined from the mean level condition (\ref{defmean}). $K=0$ for classical solitary waves (since $\eta(\infty)=0$) but for generalised solitary waves $K$ is not necessarily zero. However, via a change of definition of the mean water level (and therefore of $\depth$ and $B$), it is always possible to take $K=0$ without loss of generality. This is particularly convenient for a numerical resolution of (\ref{eq:bab1}), the mean level condition being enforced a posteriori as explained in the numerical procedure below.

The equation \eqref{eq:bab1} is a Babenko (1987) equation modified to incorporate capillarity \cite{Buffoni2000}. One advantage of Babenko-like formulations, compared to other conformal mapping based techniques \cite{Choi1999, Li2004, Milewski2010}, consists in reducing the degree of nonlinearity of the original Euler equations. This feature is advantageous, in particular, for numerical resolutions.


\section{Numerical method}\label{sec:numm}

The generalised Babenko equation \eqref{eq:bab1} is solved numerically using a Fourier collocation technique \cite{Boyd2000, Canuto2006} and a Levenberg--Marquard--Gau\ss--Newton method for solving the nonlinear algebraic system of equations resulting from the discretisation \cite{Nocedal2006}. These numerical procedures being classical, they are only briefly described here.

\subsection{Discretisation}

The conformal domain $-\infty < \alpha < \infty$ is truncated on a finite interval  $-\Lambda < \alpha\leqslant\Lambda$ and discretised at $N$ equally spaced nodes $\alpha_{j} = -\Lambda + j\Delta\alpha$, with $\Delta\alpha = 2\Lambda/N$ and $j = 1,\cdots,N$, the values $\eta_j = \eta(\alpha_{j})$ being obtained from the discrete equation \eqref{eq:bab1}.

The generalised solitary waves obtained here can be seen as the nonlinear superposition of a classical solitary wave (possibly with multiple humps) and of a periodic wave (with a single fundamental frequency). The former is thus called here the `solitary component' of the wave, while the latter is called the `periodic component'.$\Lambda$ is then chosen large enough so that the `solitary component' of the wave is zero to machine precision for $|\alpha|>\Lambda_0$ with $0<\Lambda_0<\Lambda$. $\Lambda-\Lambda_0$ is also chosen large enough so that there are several periods of the wave `periodic component'.

The domain $\alpha\in\,]-\Lambda,\Lambda]$ is considered periodic in order to apply the fast Fourier transform (FFT). $N$ is chosen large enough so that the Fourier spectrum is zero to machine precision for all the highest frequencies. Thus, the spacial and spectral truncation errors are extremely small (around machine precision). In practice, all the (pseudo) differential operations are performed in Fourier space, while all the nonlinear operations are performed in the (discrete) $\alpha$-space. With this approach, we obtain a nonlinear system of algebraic equations that must be solved numerically with an iterative procedure.

Note that the $(2\Lambda)$-periodic conformal $\alpha$-domain corresponds to a 
$(2L)$-periodic physical $x$-domain with, in general, $L\neq\Lambda$. Note also that the computational nodes are not equally spaced in the physical domain. Note finally that in the far field of a generalised solitary waves, the free surface $\eta(x)$ is $(2\ell)$-periodic while $\eta(\alpha)$ is $(2\lambda)$-periodic with, in general, $\ell\neq\lambda$.


\subsection{Nonlinear equations solver}

For pure gravity waves ($\tau=0$), the equations can be solved efficiently with the simple Petviashvili scheme \cite{Clamond2012b, Pelinovsky2004}. When $\tau\neq0$, Petviashvili's scheme fails due to the non homogeneous character of the nonlinear terms involved in the capillary terms. Several variants of the Newton method are possible alternatives to the Petviashvili scheme. The reasons for the failure of some candidates \cite{Boyd2007} were taken into account to choose a suitable method. The best results were obtained by employing the so-called Levenberg--Marquardt algorithm \cite{Nocedal2006}. We use the {\sc Matlab}\textsuperscript{\textregistered} version of this algorithm implemented in the {\sf fsolve} function of the {\em Optimisation} toolbox. Some computations were checked and confirmed with \cite{Lourakis2004} and \cite{Nielsen1999} implementations of this method.

For the computations herein below, the physical parameters are $g=d=1$, various $B$ and $\tau$ being chosen for each computation. The Levenberg--Marquardt initial relaxation parameter is set to $0.05$ and the tolerance is $10^{-12}$, the computations being performed in double precision (about sixteen digits). Most of the time, the half of the computational domain $\Lambda/\depth_0$ varies between $20$ and $40$. The number of Fourier modes is generally $N=1024$, but a larger number of nodes is sometimes used, for example with the solutions computed in the Figure~\ref{fig:multihump}.


\subsection{Initial guess}

In order to obtain multi-hump (classical and generalised) solitary waves, the initial guesses of the iterations contain the desired number of pulses. Depending on the initial guess, the algorithm may converge or not to a non-trivial solution. The solution thus obtained may not have the same number of pulses as the initial guess.

All the initial guesses we use satisfy the far field condition $\eta=0$ to machine precision for $\Lambda_0<|\alpha|\leqslant\Lambda$. Thus, the mean level condition (\ref{defmean}) is numerically fulfilled according to (\ref{defmeansolcla}) and (\ref{defmeansolgen}), and $K = 0$ for the initial guesses. For all the computations, we choose unit gravity ($g = 1$) and unit initial mean depth ($\depth_0=1$), while various surface tensions $\tau$ and initial Bernoulli constants $B_0 = c_0'^2$ are considered.

In most of the cases, our implementation of the method makes use of the natural numerical continuation in the surface tension $\tau$, as a way to improve the performance and to accelerate the convergence. Thus, for a given $B_0$, we start the algorithm with a small $\tau$ and subsequently increase $\tau$ little by little to reach the desired value.

During the computations, the mean level condition (\ref{defmean}) is {\em not\/} enforced at each iteration, as it would greatly complicate the algorithm. It is simpler to release this constraint and carry out the computations until convergence is achieved. Eventually, the physical mean depth and Bernoulli constant are computed according to their definitions given in the section \ref{sec:model}. The procedure is detailed below.


\subsection{Post-processing}\label{ssecpostpro}

An important aspect of the resolution method is the numerical definition of the averaging operations. In order to fulfil the condition (\ref{defmean}) accurately, the computational domain has to be very large: this is beyond most computer capabilities. Such a large domain is however not necessary for computations. Indeed, a classical solitary wave (and the `solitary part' of a generalised solitary wave) decaying rapidly, then there exists $\Lambda_0$ such that the solitary wave contribution is smaller than the machine precision for all $|\alpha| \geqslant \Lambda_0$ (the far field), so it does not contribute to the numerical solution in the far field. Therefore, the computational domain needs only to be large enough so that the `solitary part' of the wave is zero to machine precision for $\Lambda_0\leqslant |\alpha| \leqslant \Lambda$. Hence, in the far field, the flow is numerically $2\ell$-periodic to machine precision and can be expanded in Fourier polynomials     
\begin{eqnarray*}
  \eta(x)\ \approx\, \sum_{m=-M}^M\mathfrak{h}_m\,\ue^{\ui\/m\/\pi\/x/\ell},
\qquad |\sur{w}(x)|^2\ \approx\, \sum_{m=-M}^M\mathfrak{q}_m\,\ue^{\ui\/m\/\pi\/x/\ell},
\end{eqnarray*}
where the unknown parameters $\ell$, $\mathfrak{h}_m$ and $\mathfrak{q}_m$ are all determined by nonlinear least square minimisation \cite{Clamond1995}. According to (\ref{defmeansolgen}), the mean water level $\bar{\eta}$ is then 
\begin{equation*}
  \bar{\eta}\ =\ \frac{1}{2\,\ell}\int_{-\ell}^{\ell}\eta(x)\,\ud\/x\ =\ \mathfrak{h}_0.
\end{equation*}
In principle, $\bar{\eta}$ should be zero to the accuracy involved in the numerical procedure. If this is not the case, it means that the depth $\depth_0$ introduced {\em a priori\/} in the resolution procedure has `drifted' and should then be redefined (renormalised) in order to compute the {\em a posteriori\/} (actual) depth $\depth_\infty$ defined by
\begin{equation}\label{depthphys}
  \depth_\infty\ \equiv\ \depth_0\ +\ \bar{\eta}\ =\ \depth_0\ +\ \mathfrak{h}_0.
\end{equation}
Similarly, the {\em a posteriori\/} Bernoulli constant $B_\infty$ is then defined by
\begin{equation}\label{bernphys}
  B_\infty\ \equiv\  \frac{1}{2\,\ell}\int_{-\ell}^{\ell}\left[\,2\,g\,\eta\,+\,|\sur{w}(x)|^2\,\right]\ud\/x\ =\ 2\,g\,\mathfrak{h}_0\ +\ \mathfrak{q}_0, 
\end{equation}
and the {\em a posteriori\/} phase velocity $c_\infty'$ is thus 
\begin{equation}\label{velcphys}
  c_\infty'\ \equiv\  \sqrt{\,B_\infty\,-\,2\,g\,\bar{\eta}\,}\ =\ \sqrt{\,\mathfrak{q}_0\,}. 
\end{equation}

In practice, the choice $M = 5$ is more than sufficient because the tails of the generalised solitary waves computed here are quasi-sinusoidal (see below).


\subsection{Dimensionless parameters}

In order to characterise the waves, we introduce the dimensionless parameters
\begin{equation*}
  \Fr^2\ \equiv\ \frac{c'^{\,2}}{g\,\depth}, \qquad \Bo\ \equiv\ \frac{\tau}{g\,\depth^{\,2}}, 
\end{equation*}
$\Fr$ and $\Bo$ being, respectively, the Froude and Bond numbers. (Our definition of the Bond number is the most common in the theory water waves, but it is the reciprocal of the Bond number usually defined in fluid mechanics.)

The Froude and Bond numbers are chosen a priori to compute the solutions, i.e., the procedure described above implies initial Froude and Bond numbers $\Fr_0$ and $\Bo_0$, respectively. Once the computations have converged, we compute the a posteriori Froude and Bond numbers $\Fr_\infty$ and $\Bo_\infty$, respectively.


\section{Numerical results}\label{sec:numr}

We demonstrate here the efficiency of the algorithm computing several known and new solutions.

\begin{figure}
  \centering
  \includegraphics[width=0.99\textwidth]{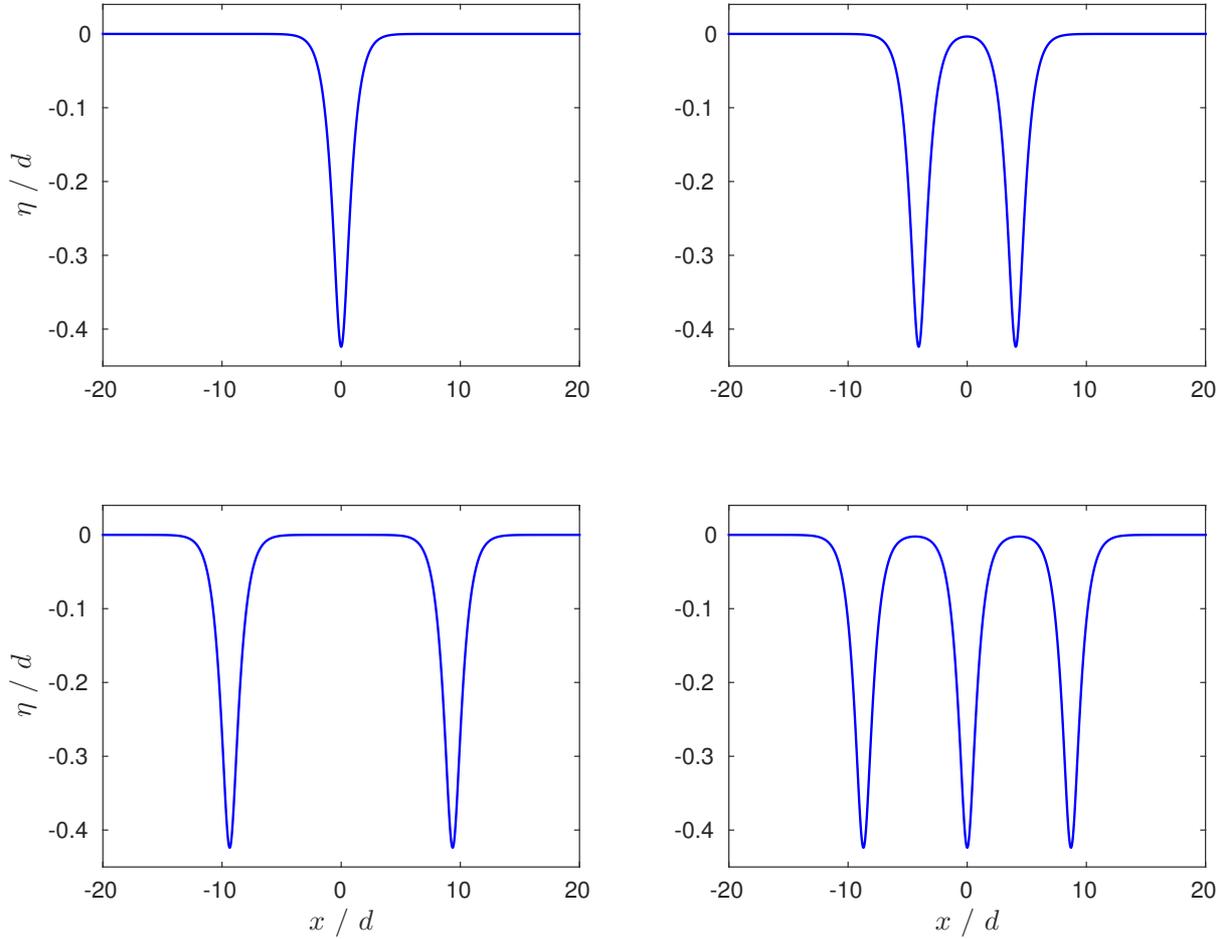}
  \caption{\small\em Examples of solitary waves of depression for $\Fr = 0.75$ and $\Bo = 0.4$.}
  \label{fig:clasol} 
\end{figure}


\subsection{Classical solitary waves}

For supercritical Bond numbers $\Bo > 1/3$ and suitable values of the Froude number $\Fr < 1$, the existence of isolated solitary waves of depression is known, theoretically and numerically, see \cite{Champneys2002} and the references therein. A first check of the code is the computation of one of these waves corresponding to the values $\Bo = 0.4$ and $\Fr = 0.75$ (Figure~\ref{fig:clasol}, upper-left). The initial guess of the procedure was chosen simply as a negative localised bump. The algorithm convergence does not seem to be sensitive to the choice of the initial guess for such simple solutions.

Another type of solution, computed by the code and shown here in Figure~\ref{fig:clasol}, consists of multi-modal solitary waves. The existence of these solutions in the KdV5 equation was shown theoretically and numerically \cite{Champneys1997}. Investigating the neighbourhood of $\Fr = 1$ and $\Bo = 1/3$, \cite{Buffoni1996a} proved, for the Euler equations, the existence of an infinite number of solitary waves.  Using a boundary integral equation technique and a different numerical procedure \cite{Hunter1983a}, multi-modal solitary waves of depression were  computed by \cite{Dias1996}. In order to obtain multi-modal solitary waves of depression, the initial guess was chosen with several negative bumps. Three examples of multi-modal solitary waves of depression are given in Figure~\ref{fig:clasol} for $\Bo = 0.4$ and $\Fr = 0.75$ (identical to the uni-modal example). This shows numerically the non-unicity of the solution for a fixed pair of parameters ($\Fr, \Bo$). Actually, solutions with any number of pulses could probably be obtained. For the same number of pulses, the distance between the pulses is also not unique (Figure~\ref{fig:clasol} upper-right and lower-left). Our numerical investigations suggest that the distance between the two pulses cannot be reduced below a certain distance. However, the distance can be increased at will. In the upper-right of Figure~\ref{fig:clasol}, the two pulses interact because $\eta(0)/\depth = -3.39\times10^{-3}$ is larger than the numerical accuracy of the computations, this case being the shortest distance we where able to compute.

All the examples in Figure~\ref{fig:clasol} have identical Froude and Bond numbers, but also identical amplitudes $\min(\eta/\depth) = -0.42396 \pm 1.3\times10^{-5}$. For these examples, the algorithm converged to the desired precision ($10^{-12}$), the different solutions being obtained only changing the initial guess. These solutions were first calculated with computational boxes of length $\Lambda/\depth = 25$ and $N = 1024$ nodes, subsequently verified with $\Lambda/\depth = 50$ and $N = 2048$.

\begin{figure}
  \centering
  \includegraphics[width=0.7\textwidth]{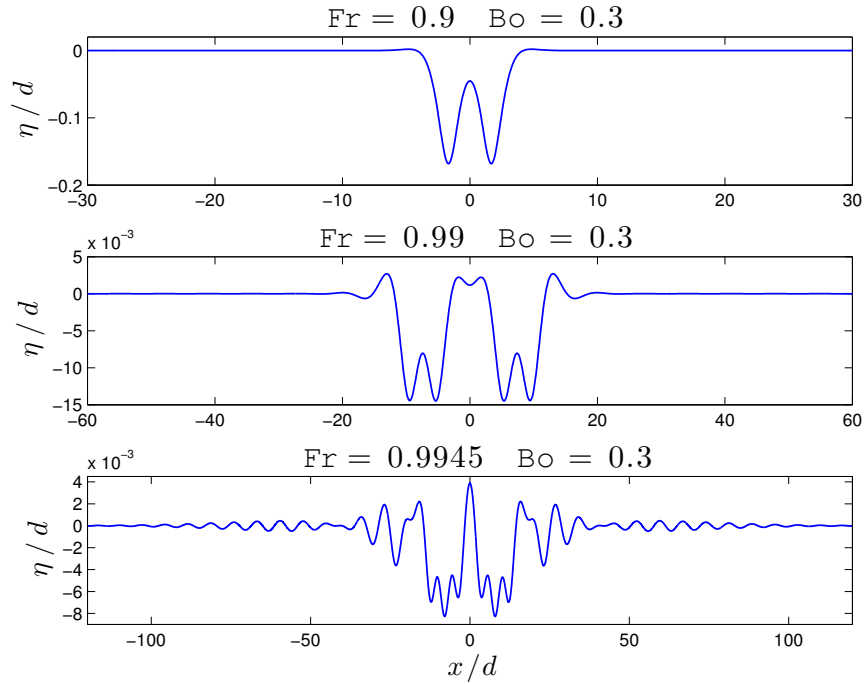}
  \caption{\small\em Examples of multi-modal solitary waves with damped oscillations.}
  \label{fig:multihump}
\end{figure}

For $\Fr < 1$ and $\Bo < 1/3$, solitary waves with tails oscillating around the rest level are obtained. A variety of multi-hump solitary waves with damped oscillatory tails is illustrated in Figure~\ref{fig:multihump} where some `exotic' waves are formed as the Froude number approaches $1$ (similar solutions have been computed by \cite{Dias1996}). These examples also illustrate the abilities of our numerical procedure. (For instance, the computation of the lower Figure~\ref{fig:multihump} requires the limit of our hardware capabilities, but more powerful computers will permit the computation of more complicated solutions.)

\begin{figure}
  \centering
  \includegraphics[width=0.99\textwidth]{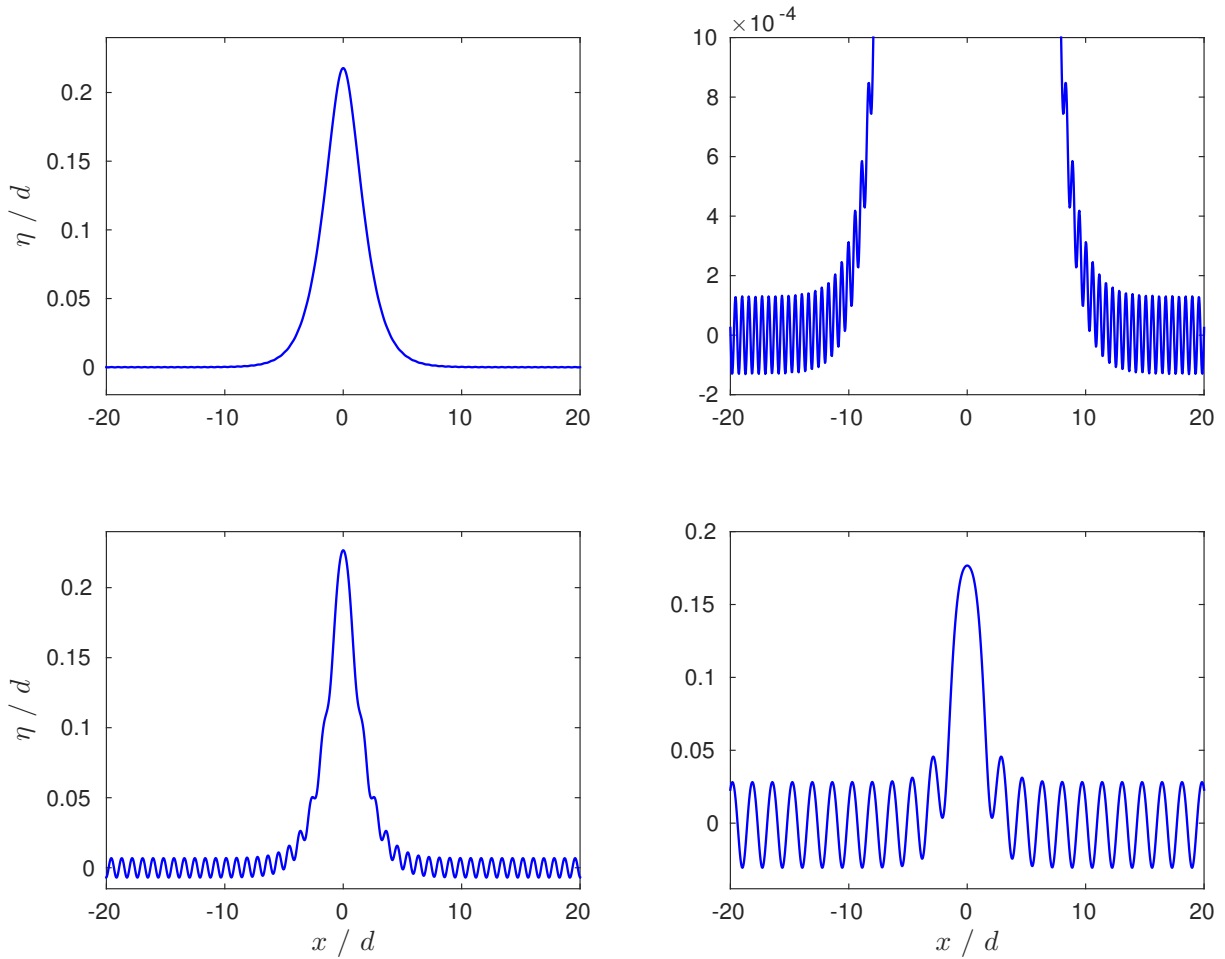}
  \caption{\small\em Examples of generalised solitary waves with $\Fr_0= 1.1$.}
  \label{fig:gensol} 
  \centerline{\scriptsize Up-left: $\Bo_0 = 0.1$, $\Bo_\infty=0.0999998$, $\Fr_\infty=1.099999$; 
  Up-right: zoom of up-left;} 
  \centerline{\scriptsize Low-left: $\Bo_0=0.15$, $\Bo_\infty=0.1499$, $\Fr_\infty=1.099$; 
  Low-right: $\Bo_0=0.25$, $\Bo_\infty=0.248$, $\Fr_\infty=1.095$.}
\end{figure}

\begin{figure}
  \centering
  \includegraphics[width=0.99\textwidth]{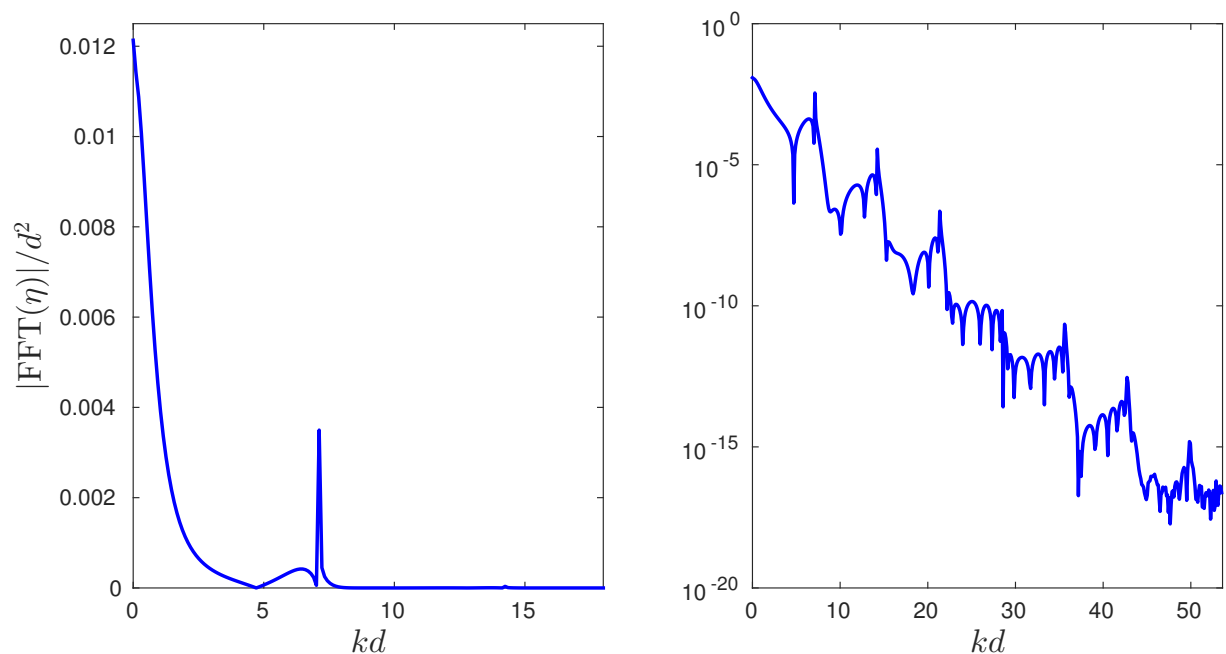}
  \caption{\small\em Fourier spectrum of Figure \ref{fig:gensol} low-left.}\label{fig:gensolfft} 
\end{figure}

\begin{figure}
  \centering
  \includegraphics[width=0.99\textwidth]{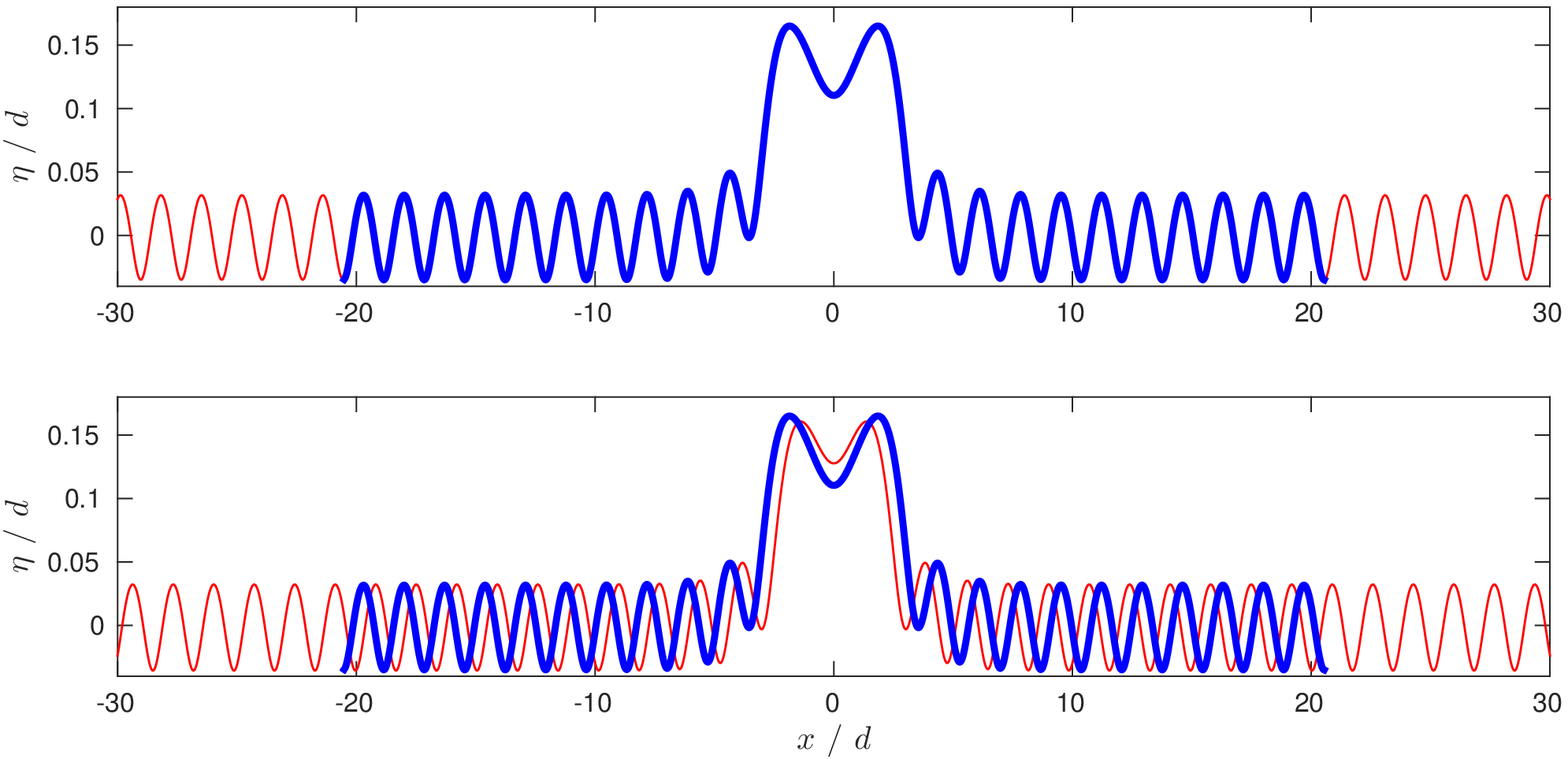}
  \caption{\small\em Influence of the computational box for $\Fr_0 = 1.1$ and $\Bo_0 = 0.25$.}
  \label{fig:gensolbox} 
  \centerline{\scriptsize Thick blue line: $\Lambda/\depth_0=20$, $\depth_\infty/d_0=1.00914$, $\Fr_\infty=1.0873$, $\Bo_\infty=0.24549$;}
  \centerline{\scriptsize Upper thin red line: $\Lambda/\depth_0=30.11$, $\depth_\infty/d_0=1.00913$, $\Fr_\infty=1.0874$, $\Bo_\infty=0.2455$;}
  \centerline{\scriptsize Lower thin red line: $\Lambda/\depth_0=30.5$, $\depth_\infty/d_0=1.0095$, $\Fr_\infty=1.0868$, $\Bo_\infty=0.24531$.}
\end{figure}


\subsection{Generalised solitary waves}

The question of existence of classical solitary waves of elevation is not solved, to our knowledge, although some references suggest a negative answer \cite{Champneys2002}. For $\Bo < 1/3$ and $\Fr > 1$, what is known is the existence of generalised solitary waves (solitary waves that are homoclinic to exponentially small amplitude oscillatory waves). Here, we focus on the numerical generation of this kind of waves.

Our experiments concern the range $\Bo < 1/3$ and $\Fr > 1$, illustrating thus the influence of the capillarity (smaller or larger values of $\Bo$) on the computations. The resulting numerical profiles are shown in Figure~\ref{fig:gensol} for the same a priori Froude number $\Fr_0 = 1.1$ and various Bond numbers. In all these cases, periodic tails emerged. The results presented here were obtained by making the continuation in the Bond number $\Bo$. As a result, starting from small values of the Bond number close to $0.1$, the numerical results suggest the emergence of generalised solitary waves as a consequence of a resonance between the solitary and the periodic waves of the same speed \cite{Beale1991}.

For $\Bo = 0.1$, the tail is not visible in the upper-left wave of Figure~\ref{fig:gensol}, but it is nevertheless present (see the zoom in Figure~\ref{fig:gensol} up-right). The amplitude of the ripples grows with the Bond number. These tails are quasi-sinusoidal, as can be seen from the Fourier transform of the free surface (Figure~\ref{fig:gensolfft}) showing than the fundamental frequency is dominant with small and exponentially decaying harmonics.

With the emergence of an oscillating tail, the mean surface level is slightly moved, requiring a renormalisation of the physical parameters. However, this drift is small in all the solution we computed, at most of a few percents in the worst cases (Figs.~\ref{fig:gensol} \& \ref{fig:gensolbox}).

The examples in Figure~\ref{fig:gensol} were obtained by continuation for the Bond number, increasing it progressively from $\Bo = 0$. If instead the solution is computed from the initial guess with a finite Bond number, a different solution may be obtained. This is illustrated for $\Fr_0 = 1.1$ and $\Bo_0 = 0.25$, see the lower-right panel in Figure~\ref{fig:gensol} and Figure~\ref{fig:gensolbox}.

For classical solitary waves, the length of the computational domain has no influence, provided that it is long enough. This is not the case for generalised solitary waves where the tails are somehow quantised by the length of the domain. Nonlinear interactions between the solitary and periodic parts of the wave inducing phase shifts, the computational domain is generally not an exact multiple of tail wavelength ($L/\ell$ and $\Lambda/\lambda$ are not integers, in general). Figure~\ref{fig:gensolbox} illustrates the influence of the computational box on the results. The solution in thick blue line was computed with $\Lambda=20\depth_0$. Recomputing the solution in a longer domain with $\Lambda = 30.11\depth_0$, we obtained the same solution (see the thin red line in the upper Figure~\ref{fig:gensolbox}). This shows that, though the computational domain is periodic, the computed solution is an accurate approximation of an aperiodic wave. However, for domains of arbitrary lengths, different solutions are generally obtained, as shown for $\Lambda=30.5\depth_0$ in the lower Figure~\ref{fig:gensolbox}. This example clearly illustrates the influence of the computational domain on the tail characteristics that can be obtained with the algorithm described in this paper. Whether or not the wavelength and amplitude of the tail can vary continuously for exact generalised solitary waves is an open question we cannot answer here.

\begin{figure}
  \centering
  \includegraphics[width=0.99\textwidth]{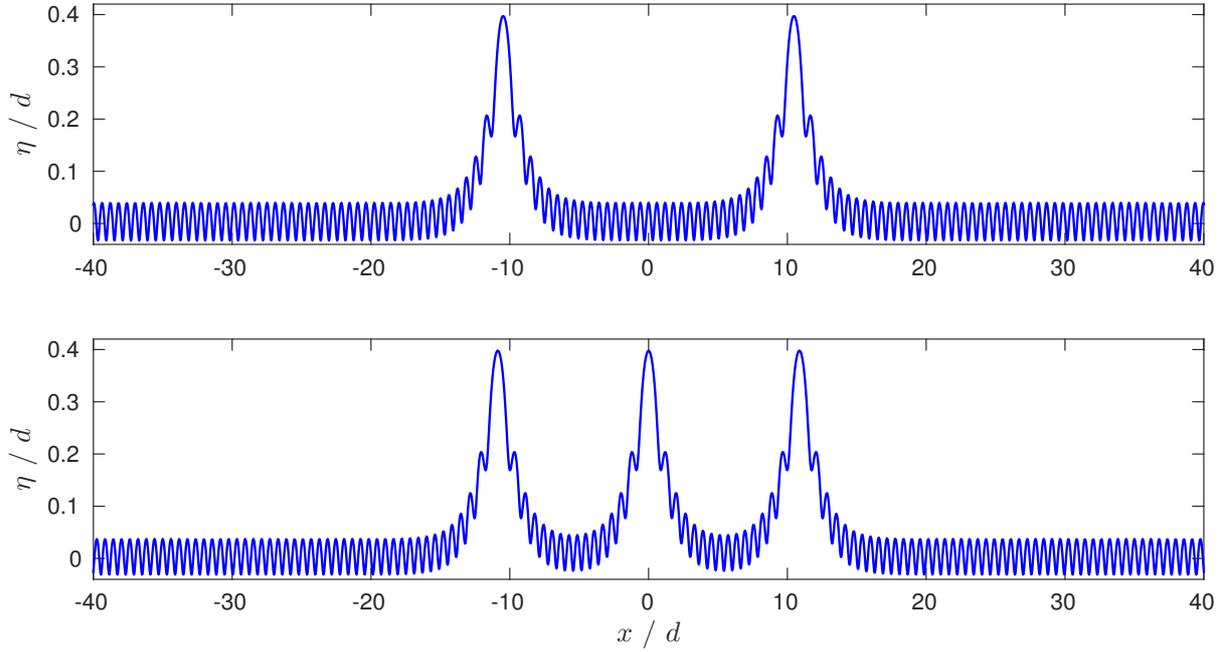}
  \caption{\small\em Multi-hump generalised solitary waves with $\Fr_0= 1.17$ and $\Bo_0=0.12$.}
  \label{fig:mhgensol} 
  \centerline{\scriptsize Upper: $\Fr_\infty=1.2466$, $\Bo_\infty=0.1181$, $\depth_\infty/\depth_0=1.008$, $\max(\eta/\depth_\infty)=0.4018$.}  
  \centerline{\scriptsize Lower: $\Fr_\infty=1.2396$, $\Bo_\infty=0.1183$, $\depth_\infty/\depth_0=1.007$, $\max(\eta/\depth_\infty)=0.4021$.}
\end{figure}

\begin{figure}
  \centering
  \includegraphics[width=0.99\textwidth]{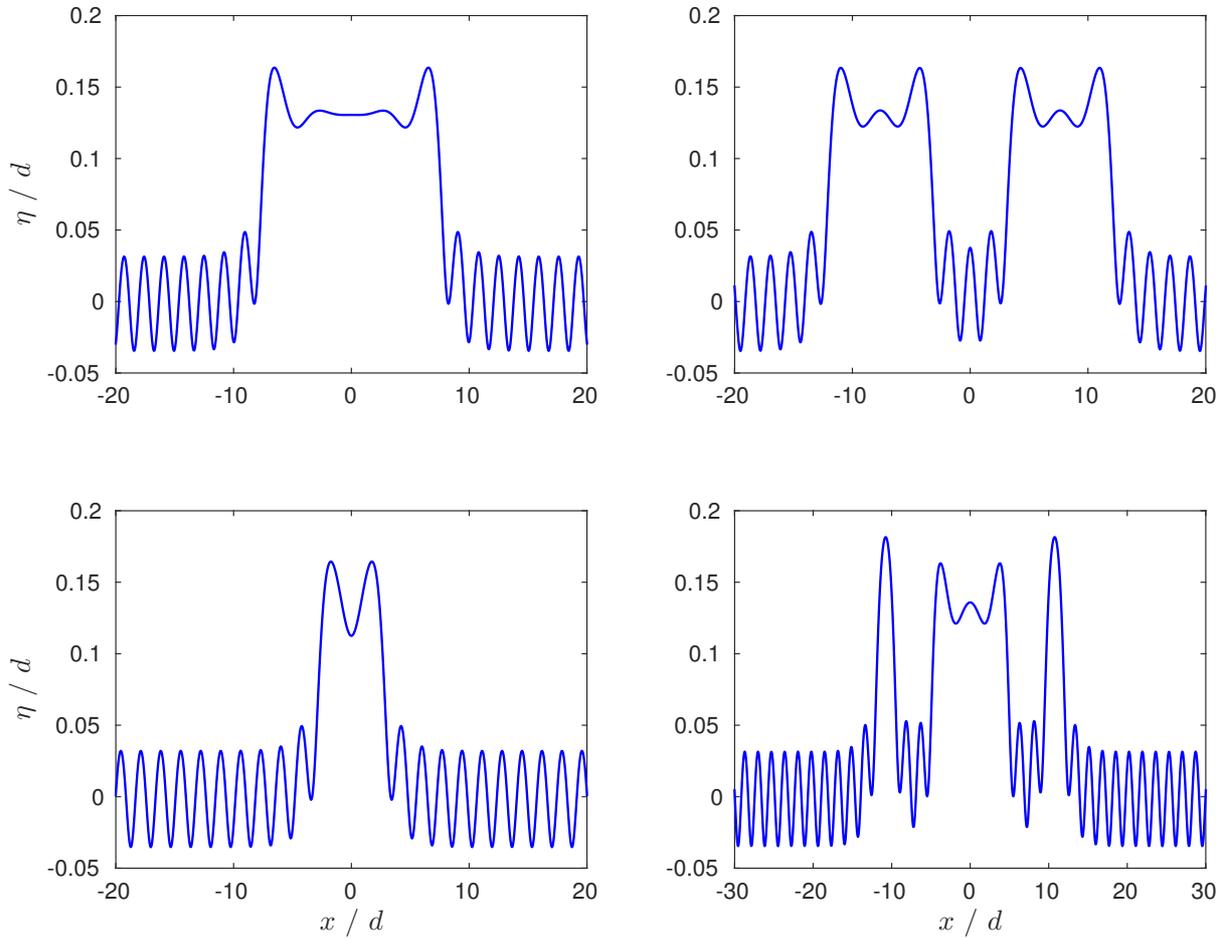}
  \caption{\small\em Unequal multi-hump generalised solitary waves with $\Fr_0=1.1$ and $\Bo_0=0.25$.}
  \label{fig:umhgensol}
  \centerline{\scriptsize Upper-left: $\Fr_\infty=1.0876$, $\Bo_\infty=0.24559$;  Upper-right: $\Fr_\infty=1.0875$, $\Bo_\infty=0.24557$;}  
  \centerline{\scriptsize Lower-left: $\Fr_\infty=1.0871$, $\Bo_\infty=0.24541$; Lower-right: $\Fr_\infty=1.0876$, $\Bo_\infty=0.24558$.}
\end{figure}


\subsection{Multi-hump generalised solitary waves}

A final type of waves computed and shown here consists of multi-hump generalised solitary waves. The existence of homoclinic connections with several loops near a resonance for a family of Hamiltonian systems has been recently analysed by \cite{Jezequel2014}. In the context of water waves, these solutions would correspond to multi-hump generalised solitary waves and we offer here some numerical evidences of their existence.

Taking a finite number of separated bumps as the initial iteration with $\Fr = 1.17$ and $\Bo = 0.12$, a generalised two-pulse capillary--gravity solitary wave is shown in the upper Figure~\ref{fig:mhgensol} while, for the same values of the parameters, a three-pulse solution is depicted in the lower Figure~\ref{fig:mhgensol}. For these two waves, the a posteriori parameters differ by less than one percent, suggesting the non-unicity of the solution. This non-unicity is to be expected since it occurs already for classical solitary waves.

Our experiments suggest that the process of adding new pulses can be continued indefinitely, provided that the length and resolution of the computational domain are gradually increased as well. This experiment provides numerical evidences that the solutions should not be unique for a given set of Froude and Bond numbers, the number of solutions being likely infinite (countable infinity in the number of pulses, and continuum of solutions varying the distance between pulses).

For a pair of classical solitary waves, we found that the two humps cannot be too close. This seems not to be the case with multi-hump generalised solitary waves because we succeeded in computed largely overlapping waves, such as in Figure~\ref{fig:umhgensol}. These solutions were all obtained for $\Fr_0=1.1$, $\Bo_0=0.25$ and $\Lambda = 30\depth_0$ (though some results are plotted on restricted domains for a better display). Except for the lower-left in Figure~\ref{fig:umhgensol}, the others very different solutions have the same (a posteriori) Froude and Bond numbers within some accuracy (their discrepancies are smaller that $0.01\%$). This is a strong numerical evidence of the non-unicity of generalised solitary waves.


\section{Discussion}\label{sec:concl}

In this study, the problem of steady capillary--gravity solitary waves was reformulated on a flat domain and the corresponding Babenko-like equation was derived. This equation was discretised with a Fourier-type pseudo-spectral method. The resulting discrete nonlinear and nonlocal equation was solved using the Levenberg--Marquardt  method. The accuracy and capabilities of the method were demonstrated via several examples.

Using this formulation we succeeded to compute, directly or by continuation in the Bond number $\Bo$, classical and generalised solitary waves. Above the critical Bond number $\Bo=1/3$, we found the classical localised solitary waves of depression which propagate with subcritical speeds $\Fr<1$, in agreement with the predictions of the KdV5 model. We also showed that various (generalised) multi-pulse solitary waves of elevation, but also of depression (localised), can be also successfully computed by our method. To our knowledge, these generalised multi-pulse solitary waves have never been computed in the context of capillary-gravity surface waves. The numerical simulations suggest the existence of an infinite number of such waves for equal Froude and Bond numbers.   

The stability of these multi-hump generalised solitary waves is a question of physical and theoretical interest. Periodic (Stokes) surface waves are known to be unstable \cite{McLean1982}. The generalised solitary waves involving periodic waves in their far field, they are likely unstable as well, but rigorous investigation have yet to be performed. 

Numerical evidences are of course not rigorous mathematical proofs, mathematical investigations would then be of great interest. Indeed, the lack of theoretical results for these multi-hump generalised solitary waves makes their physical analysis difficult. Such theoretical insights should also be beneficial for improving the computation of these waves; it would then be easier to compute other solutions, such as asymmetric generalised solitary waves (if they exist). We hope that the numerical evidences provided here will stimulate such theoretical investigations.


\subsection*{Acknowledgments}
\addcontentsline{toc}{subsection}{Acknowledgments}

D.~\textsc{Clamond} and D.~\textsc{Dutykh} acknowledge the support of the CNRS under the PEPS Inphyniti 2015 project FARA. A.~\textsc{Dur\'an} has been supported by the project MTM2014-54710-P. The authors would like to thank Professor Taras \textsc{Lakoba} (University of Vermont, USA) for stimulating discussions.

\appendix
\section{Calculation details}\label{appmath}

With the conformal mapping of non-overturning waves ($x_\alpha\geqslant0$), we have
\begin{equation}\label{exrex}
  \eta_x\left(\,1\,+\,\eta_x^{\,2}\,\right)^{-1/2}\ =\,\left[\,y_\alpha\left(\,x_\alpha^{\,2}\, + \,y_\alpha^{\,2}\,\right)^{-1/2}\,\right]_{\beta=0},
\end{equation}
thence the horizontal derivative of \eqref{exrex} yields
\begin{align}
  \frac{\ud}{\ud\/x}\!\left[\,\frac{\eta_x}{\sqrt{\,1\,+\,\eta_x^{\,2}}}\,\right]\, &=\,\left[\,\frac{x_\alpha\,y_{\alpha\alpha}\,-\,y_\alpha\,x_{\alpha\alpha}}{(\,x_\alpha^{\,2}\,+\,y_\alpha^{\,2}\,)^{3/2}}\,\right]_{\beta=0}\,=\,\left[\,\frac{z_\alpha^*\,z_{\alpha\alpha}\, - \,z_\alpha\,z_{\alpha\alpha}^*}{2\,\ui\,(\,z_\alpha\,z_\alpha^*\,)^{3/2}}\,\right]_{\beta=0}\nonumber\\ 
  &=\ \ui\left(\frac{\ud\,\sur{z}^*}{\ud\/\alpha}\right)^{\!-1}\frac{\ud}{\ud\/\alpha}\!\left[\,\frac{z_\alpha^*}{z_\alpha} \,\right]_{\beta=0}^{1/2}\,=\ \left(\frac{\ud\,\sur{z}^*}{\ud\/\alpha}\right)^{\!-1}\frac{\ud}{\ud\/\alpha}\! \left[\,\frac{\ui\,z_\alpha^*}{\,\sqrt{\,z_\alpha\,z_\alpha^*\,}\,} \,\right]_{\beta=0},
\end{align}
from which the relation \eqref{psdzs} derives at once.

\addcontentsline{toc}{section}{References}
\bibliographystyle{abbrv}
\bibliography{biblio}

\end{document}